\documentclass{ws-procs9x6}

\usepackage{psfrag}

\begin{document}

\title{PHOTOPRODUCTION OF A   $\pi\rho$ PAIR AND TRANSVERSITY  
GPDs}

\author{M. EL BEIYAD$^{*, **}$, B. PIRE$^{*}$, L. SZYMANOWSKI$^{\$}$, S.~WALLON$^{**,\$\$}$}

\address{
$^*$CPhT, \'Ecole Polytechnique, CNRS
Palaiseau, 91128, France\\
$^{**}$ LPT, Universit\'e Paris XI, CNRS, 91404 Orsay, France\\
$^{\$}$ Soltan Inst. for Nuclear Studies, Warsaw, 00-681, Poland\\
$^{\$\$}$UPMC Univ. Paris 06, Facult\'e de Physique,75252 Paris, France
}

\begin{abstract}
We demonstrate that
the chiral-odd transversity generalized parton distributions (GPDs) of the nucleon can be 
accessed  through the exclusive photoproduction process 
$\gamma + N \to \pi + \rho + N',$ in the kinematics where the meson pair has a large invariant mass 
and the final nucleon has a small transverse momentum, provided the vector meson is produced 
in a transversely polarized state. We calculate perturbatively the scattering amplitude at 
leading order in $\alpha_s.$ We build a simple model for the dominant transversity 
GPD $H_T(x,\xi, t)$ based on the concept of double distribution. 
Counting rates estimates show that the experiment
looks feasible with the real photon beam characteristics expected at 
JLab@12 GeV, in low $Q^2$ leptoproduction at Jlab@12 GeV 
and in the COMPASS experiment.
\end{abstract}


\bodymatter

\section{Chiral-odd GPDs and factorization}
\label{Sec:Int}

Transversity quark distributions in the nucleon remain among the most unknown 
leading twist hadronic observables, mainly due to their chiral-odd character which enforces their decoupling in most hard amplitudes. After the pioneering studies of Ref.~\refcite{tra}, 
much work \cite{Barone} has been devoted to the study of many channels but experimental difficulties have challenged the most promising ones.

On the other hand, tremendous progress has been recently witnessed on the QCD description of hard exclusive processes, in terms of generalized parton distributions (GPDs) describing the 3-dimensional content of hadrons. Access to the chiral-odd transversity GPDs~\cite{defDiehl}, noted  $H_T$, $E_T$, $\tilde{H}_T$, $\tilde{E}_T$, has however turned out to be even more challenging~\cite{DGP} than the usual transversity distributions: one photon or one meson electroproduction leading twist amplitudes are insensitive to transversity GPDs.  The strategy which we follow here, as initiated in Ref.~\refcite{IPST}, is to study the leading twist contribution to processes where more mesons are present in the final state. A similar strategy has also been advocated recently in Ref.~\refcite{kumano} for chiral-even GPDs. We advocate that the hard scale which allows to probe the short distance structure of the nucleon
is  $s=M_{\pi \rho}^2\, \sim |t'|$ in the fixed angle regime.
In the example developed previously~\cite{IPST}, the process under study was the high energy photo (or electro) diffractive production of two vector mesons, the hard probe being the virtual ``Pomeron'' exchange (and the hard scale was the virtuality of this pomeron), in analogy with the virtual photon exchange occuring in the deep electroproduction of a meson. 

We study here \cite{PLB} a process involving 
a transversely polarized $\rho$ meson in a 3-body final state:
\begin{equation}
\gamma + N \rightarrow \pi + \rho_T + N'\,.
\label{process}
\end{equation}
It is a priori sensitive to chiral-odd GPDs due to the chiral-odd character of the leading twist distribution amplitude (DA) of $\rho_T$.  The estimated rate depends of course much on the magnitude of the chiral-odd GPDs. Not much is known about them, but model calculations have been developed 
in Refs.~\refcite{IPST, Sco, Pasq, othermodels} and  a few moments have been computed on the lattice~\cite{lattice}.
To factorize the amplitude of this process we use  the now classical proof of the factorization of exclusive scattering at fixed angle and large energy~\cite{LB}. The amplitude for the process $\gamma + \pi \rightarrow \pi + \rho $ is written as the convolution of mesonic DAs  and a hard scattering subprocess amplitude $\gamma +( q + \bar q) \rightarrow (q + \bar q) + (q + \bar q) $ with the meson states replaced by collinear quark-antiquark pairs. 
We then extract from the factorization procedure of the deeply virtual Compton scattering amplitude near the forward region the right to replace one entering meson DA by a $N \to N'$ GPD, and thus get Fig.~1. 
The needed skewness parameter $\xi$  is written in terms of the meson pair squared invariant mass
$M^2_{\pi\rho}$ as
\begin{equation}
\label{skewedness}
\xi = \frac{\tau}{2-\tau} ~~~~,~~~~\tau =
\frac{M^2_{\pi\rho}}{S_{\gamma N}-M^2}\,.
\end{equation}

\begin{figure}
\begin{center}
\psfrag{z}{\begin{small} $z$ \end{small}}
\psfrag{zb}{\raisebox{0cm}{ \begin{small}$\bar{z}$\end{small}} }
\psfrag{gamma}{\raisebox{+.1cm}{ $\,\gamma$} }
\psfrag{pi}{$\,\pi$}
\psfrag{rho}{$\,\rho$}
\psfrag{TH}{\hspace{-0.2cm} $T_H$}
\psfrag{tp}{\raisebox{.5cm}{\begin{small}     $t'$       \end{small}}}
\psfrag{s}{\hspace{.6cm}\begin{small}$s$ \end{small}}
\psfrag{Phi}{ \hspace{-0.3cm} $\phi$}
\psfrag{piplus}{$\,\pi^+$}
\psfrag{rhoT}{$\,\rho^0_T$}
\psfrag{M}{\hspace{-0.3cm} \begin{small} $M^2_{\pi \rho}$ \end{small}}
\psfrag{x1}{\hspace{-0.5cm} \begin{small}  $x+\xi $  \end{small}}
\psfrag{x2}{ \hspace{-0.2cm}\begin{small}  $x-\xi $ \end{small}}
\psfrag{N}{ \hspace{-0.4cm} $N$}
\psfrag{GPD}{ \hspace{-0.6cm}  $GPDs$}
\psfrag{Np}{$N'$}
\psfrag{t}{ \raisebox{-.1cm}{ \hspace{-0.5cm} \begin{small}  $t$  \end{small} }}
\psfig{file=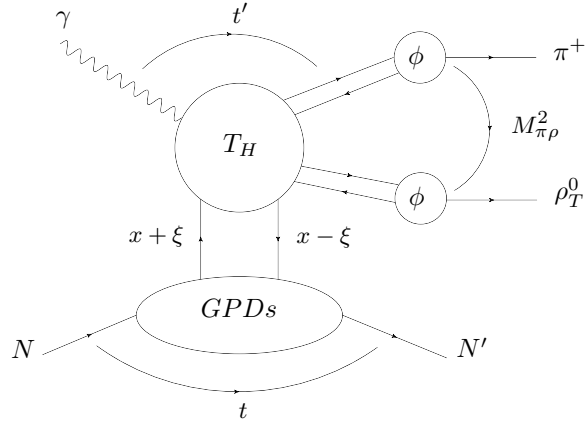,width=3in}
\caption{ Factorization of the amplitude  for $\gamma + N \rightarrow \pi + \rho +N'$ at large $M_{\pi\rho}^2$.}
\label{aba:fig1}
\end{center}
\end{figure}

Indeed the same collinear factorization property bases the validity of the leading twist approximation which either replaces the meson wave function by its DA or the $N \to N'$ transition by nucleon GPDs. A slight difference is that light cone fractions ($z, 1- z$) leaving the DA are positive, while the corresponding fractions ($x+\xi,\xi-x$) may be positive or negative in the case of the GPD. Our  Born order calculation  shows that this difference does not ruin the factorization property.

In order for the factorization of a partonic amplitude to be valid, and the
leading twist calculation to be sufficient, one should avoid the dangerous
kinematical regions where a small momentum transfer is exchanged in the
upper blob, namely small $t' =(p_\pi -p_\gamma)^2$ or small
$u'=(p_\rho-p_\gamma)^2$, and the resonance regions for each  of the
invariant squared masses $(p_\pi +p_{N'})^2$, \mbox{$(p_\rho +p_{N'})^2,$} $(p_\pi +p_\rho)^2\,.$

Let us finally stress that our discussion applies as well to the case of electroproduction where a moderate virtuality of the initial photon may help to access the perturbative domain with a lower value of the hard scale $M_{\pi \rho}$.

\section{The scattering amplitude}
\label{Sec:scattering}
The scattering amplitude of the process (\ref{process}) is written in the factorized form:
\begin{equation}
\label{AmplitudeFactorized}
\mathcal{A}(t,M^2_{\pi\rho},u')  = \int_{-1}^1dx\int_0^1dv\int_0^1dz\ T^q(x,v,z) \, H^{q}_T(x,\xi,t)\Phi_\pi(z)\Phi_\bot(v)\,,
\end{equation}
where
$T^q$ is the hard part of the amplitude and
the transversity GPD of a parton $q$  in the nucleon target 
which dominates at small momentum transfer is defined~\cite{defDiehl} by
\begin{eqnarray}
&&\langle N'(p_2),\lambda'|\bar{q}\left(-\frac{y}{2}\right)\sigma^{+j}\gamma^5 q \left(\frac{y}{2}\right)|N(p_1),\lambda \rangle  \nonumber \\
&=& \bar{u}(p',\lambda')\sigma^{+j}\gamma^5u(p,\lambda)\int_{-1}^1dx\ e^{-\frac{i}{2}x(p_1^++p_2^+)y^-}H_T^q\,,
\end{eqnarray}
where $\lambda$ and $\lambda'$ are the light-cone helicities of the nucleon $N$ and $N'$.
The chiral-odd  DA for the  transversely polarized meson vector $\rho_T$,  is defined, in leading twist 2, by the matrix element 
\begin{eqnarray}
\langle 0|\bar{u}(0)\sigma^{\mu\nu}u(x)|\rho^0_T(p,\epsilon_\pm) \rangle = \frac{i}{\sqrt{2}} (\epsilon^\mu_{\pm}(p) p^\nu - \epsilon^\nu_{\pm}(p) p^\mu)f_\rho^\bot\int_0^1du\ e^{-iup\cdot x}\ \phi_\bot(u)\,,
\end{eqnarray}
where $\epsilon^\mu_{\pm}(p_\rho)$ is the $\rho$-meson transverse polarization and with $f_\rho^\bot$ = 160 MeV. Two classes  of 
Feynman diagrams  without  (Fig.~2) and  with (Fig.~3) a 3-gluon vertex describe this process.

\begin{figure}
\begin{center}
\psfrag{fpi}{$\,\phi_\pi$}
\psfrag{fro}{$\,\phi_\rho$}
\psfrag{z}{\begin{small} $z$ \end{small}}
\psfrag{zb}{\raisebox{-.1cm}{ \begin{small}$\hspace{-.3cm}-\bar{z}$\end{small}} }
\psfrag{v}{\begin{small} $v$ \end{small}}
\psfrag{vb}{\raisebox{-.1cm}{ \begin{small}$\hspace{-.3cm}-\bar{v}$\end{small}} }
\psfrag{gamma}{$\,\gamma$}
\psfrag{pi}{$\,\pi^+$}
\psfrag{rho}{$\,\rho^0_T$}
\psfrag{N}{$N$}
\psfrag{Np}{$\,N'$}
\psfrag{H}{\hspace{-0.2cm} $H^{ud}_T(x,\xi,t_{min})$}
\psfrag{hard}{\hspace{-0.2cm} $H^{ud}_T(x,\xi,t_{min})$}
\psfrag{p1}{\begin{small}     $p_1$       \end{small}}
\psfrag{p2}{\begin{small} $p_2$ \end{small}}
\psfrag{p1p}{\hspace{-0.6cm}  \begin{small}  $p_1'=(x+\xi) p$  \end{small}}
\psfrag{p2p}{\hspace{-0.2cm} \begin{small}  $p_2'=(x-\xi) p$ \end{small}}
\psfrag{q}{\begin{small}     $q$       \end{small}}
\psfrag{ppi}{\begin{small} $p_\pi$\end{small}}
\psfrag{prho}{\begin{small} $p_\rho$\end{small}}
\psfig{file=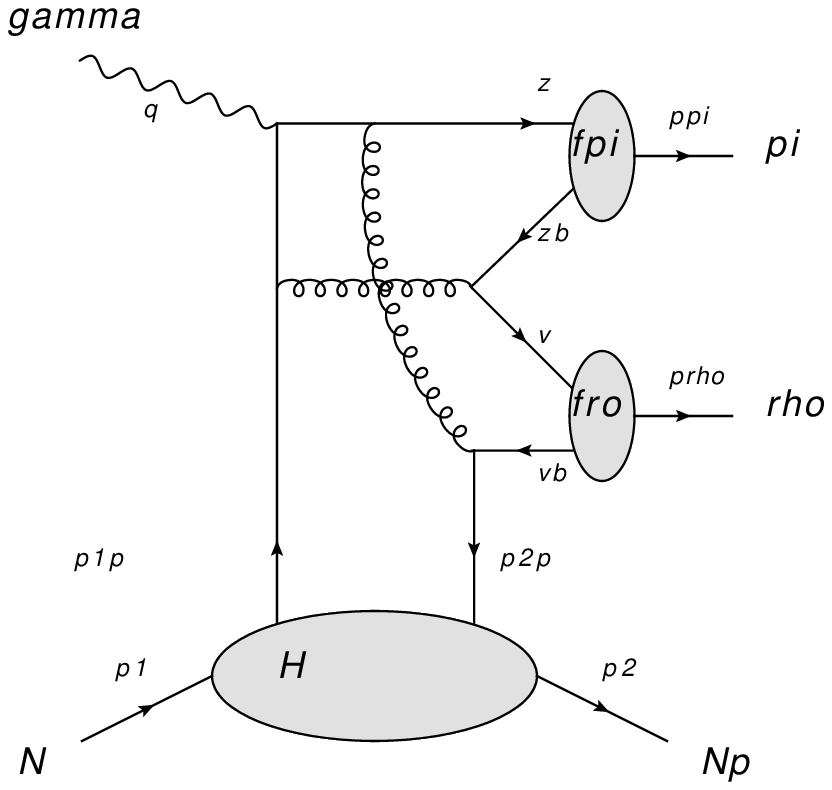,width=3in}
\caption{Sample diagram without a 3-gluon vertex.}
\label{aba:fig2}
\end{center}
\end{figure}

\begin{figure}
\begin{center}
\psfrag{fpi}{$\,\phi_\pi$}
\psfrag{fro}{$\,\phi_\rho$}
\psfrag{z}{\begin{small} $z$ \end{small}}
\psfrag{zb}{\raisebox{-.1cm}{ \begin{small}$\hspace{-.3cm}-\bar{z}$\end{small}} }
\psfrag{v}{\begin{small} $v$ \end{small}}
\psfrag{vb}{\raisebox{-.1cm}{ \begin{small}$\hspace{-.3cm}-\bar{v}$\end{small}} }
\psfrag{gamma}{$\,\gamma$}
\psfrag{pi}{$\,\pi^+$}
\psfrag{rho}{$\,\rho^0_T$}
\psfrag{N}{$N$}
\psfrag{Np}{$\,N'$}
\psfrag{H}{\hspace{-0.2cm} $H^{ud}_T(x,\xi,t_{min})$}
\psfrag{p1}{\begin{small}     $p_1$       \end{small}}
\psfrag{p2}{\begin{small} $p_2$ \end{small}}
\psfrag{p1p}{\hspace{-0.6cm}  \begin{small}  $p_1'=(x+\xi) p$  \end{small}}
\psfrag{p2p}{\hspace{-0.2cm} \begin{small}  $p_2'=(x-\xi) p$ \end{small}}
\psfrag{q}{\begin{small}     $q$       \end{small}}
\psfrag{ppi}{\begin{small} $p_\pi$\end{small}}
\psfrag{prho}{\begin{small} $p_\rho$\end{small}}
\psfig{file=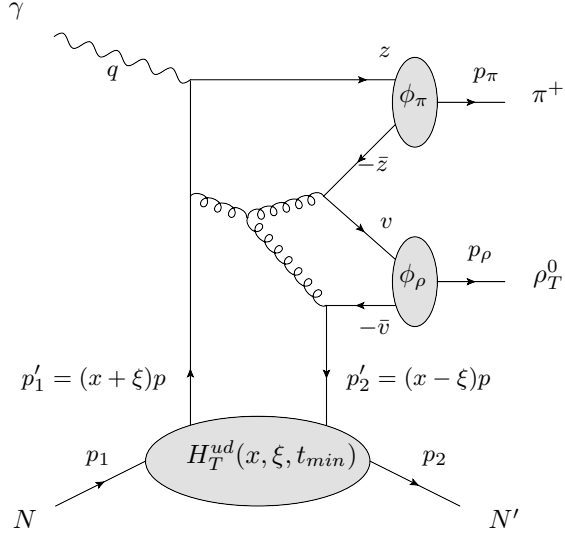,width=3in}
\caption{Sample diagram with a 3 gluon vertex.}
\label{aba:fig3}
\end{center}
\end{figure}

The scattering amplitude gets both a real and an imaginary part. Integrations over $v$ and $z$ have been done analytically whereas numerical methods are used for the integration over $x$.

\section{Results}
\label{sec:results}

Various observables can be calculated with this amplitude. We stress that even the unpolarized differential cross-section $\frac{d\sigma}{dt \,du' \, dM^2_{\pi\rho}}$
is sensitive to the transversity GPD.
To estimate the rates, we modelize the dominant transversity GPD $H_T^q(x,\xi,t)$ ($q=u,\ d$) in
terms of double distributions 
\begin{equation}
\label{DDdef}
H_T^q(x,\xi,t=0) = \int_\Omega d\beta\, d\alpha\ \delta(\beta+\xi\alpha-x)f_T^q(\beta,\alpha,t=0)
\,,
\end{equation}
with $\Omega=\{|\alpha|^2 + |\beta|^2 \leq 0 \}$ and 
where $f_T^q$ is the quark transversity double distribution written as 
\begin{equation}
\label{DD}
f_T^q(\beta,\alpha,t=0) = \Pi(\beta,\alpha)\,\delta \, q(\beta)\Theta(\beta) -
\Pi(-\beta,\alpha)\,\delta \bar{q}(-\beta)\,\Theta(-\beta)\,,
\end{equation}
where $ \Pi(\beta,\alpha) = \frac{3}{4}\frac{(1-\beta)^2-\alpha^2}{(1-\beta)^3}$ is a profile
function and $\delta q$, $\delta \bar{q}$ are the quark and antiquark transversity parton
distribution functions  of Ref.~\refcite{Anselmino}. The $t$-dependence
of these chiral-odd GPDs - and its Fourier transform in terms of the transverse
localization of quarks in the proton \cite{impact} - is very interesting but completely unknown.
We
 describe it in a simplistic way using a dipole form factor: 
\begin{equation}
\label{t-dep}
H^q_T(x,\xi,t) = H^q_T(x,\xi,t=0)\times \frac{C^2}{(t  - C)^2} \qquad (C=0.71~{\rm GeV}^2)\,.
\end{equation}
In Fig.~4,  we show the $M^2_{\pi\rho}$ dependence of the
differential cross section $\displaystyle d \sigma/d M_{\pi \rho}^2.$ for JLab kinematics ($s_{\gamma N} = 20 \,{\rm GeV}^2$). On the same plot, we show the similar cross section for the case where the $\rho$ meson is longitudinally polarized. Note that in this case (which will be discussed in detail in a forthcoming publication), the GPDs which contribute are the usual chiral even ones, parametrized here through the usual double distribution ansatz.

\psfrag{dsigdM2TLS20}{\raisebox{.5cm}{{\hspace{-.6cm}$\;\;\;\;\;\;\displaystyle  
\frac{d \sigma_{T,L}}{d M_{\pi\rho}^2}$ \hspace{0cm}{ (nb.GeV 
$^{-2}$)}}}}
\psfrag{M2}{\raisebox{-.7cm}{{\hspace{-2.5cm}$M_{\pi\rho}^2$ 
\hspace{0.1cm}(GeV$^2$)}}}
\vskip.2in
\begin{figure}[!h]
\centerline{\includegraphics[width=8cm]{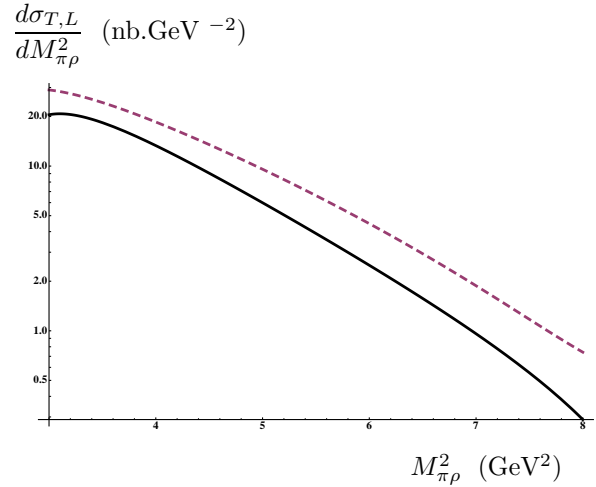}}
\vspace{.6cm}
\caption{
Differential cross section 
$\frac{d \sigma}{d M^2_{\pi \rho}}$ at
$s_{\gamma N} =20$ GeV$^2$ for the process
$\gamma p \to \pi^+ \rho^0 n$ for a transversely polarized �(full curve)
and a longitudinally polarized (dashed curve)�$\;\rho\;$�meson. The cuts are
discussed in detail in Ref.~\refcite{PLB}.
}
\label{resultM2rho0TL}
\end{figure}

Rate estimates \cite{PLB} show that expected luminosities at JLab (after the 12 GeV energy increase) and at the Compass experiment at CERN are sufficient to gather a sample of interesting events allowing this physics to be experimentally tested. 

Note that in the case of leptoproduction, the virtuality $Q^2$ of the exchanged photon plays no crucial role in our process, and the virtual photoproduction cross section is almost $Q^2$-independent if we choose to select events in a
sufficiently narrow $Q^2-$window ($.02<Q^2<1 $ GeV$^2$), which is legitimate since the
effective photon flux is strongly peaked at very low values of $Q^2$.  This  procedure applied to muoproduction at Compass
yields a rate sufficient to get an estimate of the transversity GPDs in the region of small $\xi$ ($\sim
0.01$).

Target transverse spin asymetries need to consider nucleon helicity flip amplitudes, i.e. the effect of $E_T(x,\xi,t)$ and/or  $\tilde E_T(x,\xi,t)$ GPDs. This will be studied later on.

In conclusion, we expect the process discussed here to be observable in  two quite different
energy ranges, which  should give complementary information on the chiral-odd transversity GPDs:
the large $\xi$ region may be scrutinized at JLab and the smaller $\xi$ region may be
studied at COMPASS. Let us stress that our study is built on known
leading twist factorization theorems, i.e. the scattering amplitude involves only leading twist
 non-perturbative components. This stays in contrast with other attempts to access transversity GPDs
\cite{liuti}.

\section*{Acknowledgments}

This work is partly supported by  the ANR-06-JCJC-0084 and the Polish Grant N202 249235.

\end{document}